\documentclass[journal]{IEEEtran}

\ifCLASSINFOpdf

\else

\fi

\usepackage{graphicx}
\usepackage{amsthm,amssymb,amsfonts}
\usepackage{epsfig} 
\usepackage{epstopdf}
\usepackage{amsmath}
\usepackage{amsthm}
\usepackage{comment}
\usepackage{url}
\usepackage{algorithm}
\usepackage{algorithmic}
\usepackage{subfigure}
\usepackage{booktabs}

\usepackage{amsfonts}
\usepackage{bm}
\usepackage{color}
\usepackage{xcolor}
\usepackage{cite}
\usepackage{stfloats}
\usepackage{chemarrow}
\usepackage{extarrows}
\usepackage{setspace}

\usepackage{microtype}
\usepackage[american]{babel}

\usepackage[explicit]{titlesec}
\titlespacing{\section}{0pt}{1.2ex plus .0ex minus .0ex}{.3ex plus .0ex}
\titlespacing{\subsection}{0pt}{1.2ex plus .0ex minus .0ex}{.3ex plus .0ex}

\allowdisplaybreaks

\newboolean{showcomments}
\setboolean{showcomments}{true}
\newcommand{\wang}[1]{\ifthenelse{\boolean{showcomments}}
	{ \textcolor[rgb]{1,0,1}{(ZW:  #1)}}{}}
\newcommand{\fliu}[1]{\ifthenelse{\boolean{showcomments}}
	{ \textcolor{blue}{(FL:  #1)}}{}}
\newcommand{\ychen}[1]{\ifthenelse{\boolean{showcomments}}
	{ \textcolor{green}{(ZP:  #1)}}{}}
\newcommand{\slow}[1]{\ifthenelse{\boolean{showcomments}}
	{ \textcolor{blue}{(SL:  #1)}}{}}

\theoremstyle{definition}

\theoremstyle{definition}

\def\BibTeX{{\rm B\kern-.05em{\sc i\kern-.025em b}\kern-.08em
		T\kern-.1667em\lower.7ex\hbox{E}\kern-.125emX}}

\title{\LARGE Equivalent Inertia Provided by Droop Control\\ of Fast Frequency Regulation Resources}

\begin{document}
\setstretch{1}
\author{
	Ye~Liu,~\IEEEmembership{Student Member, IEEE}, 
	and Chen~Shen,~\IEEEmembership{Senior Member, IEEE},

		\thanks{Ye Liu and Chen Shen are with the State Key Laboratory of Power Systems, Department of Electrical Engineering, Tsinghua University, Beijing 100084, China (e-mail: liuye18@mails.tsinghua.edu.cn; shenchen@mail.tsinghua.edu.cn). (\textit{Corresponding author: Chen Shen}.)}
		
			
}

\markboth{Journal of \LaTeX\ Class Files,~Vol.~xx, No.~xx, xx~xxxx}%
{Shell \MakeLowercase{\textit{et al.}}: Bare Demo of IEEEtran.cls for IEEE Journals}

\maketitle

\begin{abstract}
    Conventional inertia control strategies (i.e., the VSG-based and PD-based strategies) for the fast frequency regulation resources (FFRs) might lead to system oscillations due to their emulations of generator dynamics. The only-droop-based inertia control strategy is a feasible idea to address this issue. This letter quantitatively investigates the equivalent inertia provided by the droop control of the FFRs. First, an equivalent-scenario-based method is proposed to evaluate the equivalent inertia provided by the droop control, which shows that the droop control with a constant droop coefficient provides time-variant equivalent inertia. Then, the time-variant-droop-based equivalent inertia control (VDIC) strategy is proposed for the FFRs, which can provide the required constant inertia support without introducing generator-like dynamics. A case study verifies the effectiveness of the VDIC strategy.   
\end{abstract}

\begin{IEEEkeywords}
	Equivalent inertia support, droop control, fast frequency regulation resources, time-variant droop coefficient.
\end{IEEEkeywords}


\section{Introduction}

In a power system with high penetration of renewables and power electronics, inertia shortage is an issue of common concern. To address this issue, when a power imbalance occurs, the fast frequency regulation resources (FFRs, e.g., energy storage systems, renewable energy generation systems, and HVDC systems) are required to provide equivalent/virtual inertia support for the low-inertia power grid \cite{milano2018foundations}. The existing inertia control strategies for the FFRs can be mainly divided into two categories, i.e., the virtual synchronous generator (VSG)-based strategies \cite{chen2020modelling} and the proportional derivative (PD)-based strategies \cite{shang2021fast}. However, the above strategies essentially emulate the swing dynamics of conventional generators to provide virtual inertia support. These control dynamics might lead to system oscillations in case of improper parameter settings \cite{du2019power}.


The active power-frequency (P-f) droop control, as a simple and effective approach without control dynamics, is widely used in the FFRs to provide primary frequency support \cite{meng2018generalized}. Several studies imply that the droop control can also provide inertia support for the power system \cite{van2015droop}. Hence, if an only-droop-based inertia control strategy can be reasonably designed, the system inertia requirements will be met without introducing generator-like control dynamics. Nevertheless, there is no quantitative research answering \textit{how much equivalent inertia can be provided by the droop control}, which makes it challenging to design a droop control guided by the system inertia requirements.

To fill this gap, this letter introduces a new idea to investigate the equivalent inertia provided by the droop control of the FFRs. The contributions of this letter are twofold. First, a general equivalent-scenario-based method is proposed to evaluate the equivalent inertia provided by the droop control of the FFRs. Through this method, the quantitative relationship between the droop coefficient and the equivalent inertia support is derived, which shows that the droop control with a constant droop coefficient provides time-variant equivalent inertia. Second, a novel time-variant-droop-based equivalent inertia control (VDIC) strategy is proposed for the FFRs, which can provide the required constant equivalent inertia support without introducing generator-like dynamics.

\section{Methodology}

\subsection{Equivalent-Scenario-Based Equivalent Inertia Evaluation}

Considering a power system with multiple generators and multiple FFRs, we first make the following assumptions for the system model.

\textbf{Assumption 1:} 1) We ignore the differences among the frequencies of buses in the transient process. 2) We do not consider the damping items and the primary frequency regulation items in the swing equations of generators.

The second assumption above is reasonable for two reasons: 1) The following equivalent inertia evaluation results are conservative with this assumption. 2) The governors and turbines of the generators usually have relatively slow dynamics; hence, the primary frequency regulation items have little effect on the post-fault system inertia response. Under Assumption 1, the aggregated swing equation of the multi-generator power system is: 
\begin{align}
T_J \frac{d \omega}{dt} = \Delta P_f
\end{align}
where $\omega$ is the system frequency deviation from the nominal value, $d\omega/dt$ is the rate of change of frequency (RoCoF). $T_J = (\sum_{i\in\mathcal{G}} S_i T_{Ji})/S_{sys}$ is the total inertia constant of the system, where $\mathcal{G}$ denotes the set of generators, $S_i$ and $T_{Ji}$ are the nominal power and inertia constant of generator $i$, and $S_{sys}$ is the nominal power of the system. $\Delta P_f = P_m - P_e$ is the power imbalance caused by a fault, where $P_m$ and $P_e$ are the total mechanical power and total electromagnetic power of generators. $\Delta P_f$ is negative for power shortage while positive for power redundancy.

According to the above model, we introduce the equivalent-scenario-based equivalent inertia evaluation method. We set the following two scenarios:

\textit{Scenario 1}. The actual inertia constant of the system increases by $\Delta T_J$, while the output power of the FFRs remains unchanged. Let $\omega_1$ denotes the system frequency deviation in Scenario 1, we have:
\begin{align}
\label{scen_1}
\frac{d \omega_1}{dt} = \frac{\Delta P_f}{T_J+\Delta T_J}
\end{align}

\textit{Scenario 2}. The total output power of the FFRs changes by $\Delta P_r$ when the power imbalance occurs, while the actual inertia constant of the system remains unchanged. Let $\omega_2$ denotes the system frequency deviation in Scenario 2, we have:
\begin{align}
\label{scen_2}
\frac{d \omega_2}{dt} = \frac{\Delta P_f+\Delta P_r}{T_J}
\end{align}

The main idea of the equivalent-scenario-based method is making these two scenarios equivalent to evaluate the equivalent inertia provided by the FFRs. Since the inertia mainly affects the RoCoF, we let $d\omega_1/dt = d\omega_2/dt := d\omega_e/dt$ to make these two scenarios equivalent, where $\omega_e$ denotes the frequency deviation of the equivalent scenarios. Let $t=0$ be the time when the power imbalance occurs, we have $\omega_1(0) = \omega_2(0) = \omega_e(0)=0$; then, the equalities of the RoCoFs yield $\omega_1(t) = \omega_2(t) = \omega_e(t), \forall t\ge 0$. Combining \eqref{scen_1} and \eqref{scen_2}, we have:
\begin{align}
\label{equi_scen}
\frac{d \omega_e}{dt} = \frac{\Delta P_f}{T_J+\Delta T_J} = \frac{\Delta P_f+\Delta P_r}{T_J}
\end{align}
From \eqref{equi_scen}, we can derive:
\begin{align}
\label{gene_rep}
\Delta T_J = -\frac{T_J \Delta P_r}{\Delta P_f+\Delta P_r}
\end{align}
Equation \eqref{gene_rep} is the general quantitative relationship between the total power regulation of the FFRs and the inertia constant increment of the system. According to \eqref{equi_scen} and \eqref{gene_rep}, \textit{the post-fault fast power regulation} $\Delta P_r$ \textit{actually provides equivalent inertia support} $\Delta T_J$ \textit{for the power system, i.e., the essence of equivalent inertia is regulating output power rapidly to offset the power imbalance}. We redefine $\Delta T_J$ as the equivalent inertia coefficient.

Specifically, considering the droop control of the FFRs, the power regulation of the FFRs can be represented as: 
\begin{align}
\label{droop_des}
\Delta P_r = -k^r \omega_e
\end{align}
where $k^r = \sum_{i\in\mathcal{R}} k^r_i$ is the total droop coefficient of the FFRs, $\mathcal{R}$ denotes the set of FFRs, and $k_i^r>0$ is the droop coefficient of FFR $i$. Substitute \eqref{droop_des} into \eqref{equi_scen}, we have:
\begin{align}
\label{droop_rela}
\frac{d \omega_e}{dt} = \frac{\Delta P_f}{T_J+\Delta T_J} = \frac{\Delta P_f-k^r \omega_e}{T_J}
\end{align}  
To evaluate the equivalent inertia provided by the droop control, we let the droop coefficient $k^r$ be a constant and the equivalent inertia coefficient $\Delta T_J$ be a variable. Since $k^r$ is a constant, according to the equality between the first and third items in \eqref{droop_rela}, we can derive $\omega_e(t)$ based on the Laplace transformation:
\begin{align}
\label{omega_t}
\omega_e (t) = \frac{\Delta P_f}{k^r} (1-e^{-\frac{k^r}{T_J} t})
\end{align} 
Substitute \eqref{omega_t} into the second equality in \eqref{droop_rela}, we can derive the time-domain expression of the equivalent inertia coefficient:
\begin{align}
\label{equ_iner}
\Delta T_J (t) = T_J (e^{\frac{k^r}{T_J} t}-1)
\end{align}

\textbf{Remark 1:} First, equation \eqref{equ_iner} shows the quantitative relationship between the droop coefficient and the equivalent inertia coefficient, which illustrates that the droop control with a constant droop coefficient provides time-variant equivalent inertia support. When $t \rightarrow 0$, $\Delta T_J (t) \rightarrow 0$; when $t \rightarrow +\infty$, $\Delta T_J (t) \rightarrow +\infty$. Thus, the constant droop control provides little equivalent inertia support within a short time after the power imbalance. Inspired by this time-variant characteristic, to provide the required constant equivalent inertia support, we design a time-variant-droop-based inertia control strategy for the FFRs, as shown in Section II.B. Second, equation \eqref{equ_iner} also shows that the equivalent inertia support from the droop control of the FFRs is independent of the power imbalance $\Delta P_f$. Moreover, the equivalent inertia evaluation method presented in this subsection is general and can be extended to various power-frequency control strategies of the FFRs. 
 
\subsection{Time-Variant-Droop-Based Equivalent Inertia Control}

In this subsection, we design the time-variant-droop-based equivalent inertia control (VDIC) strategy for the FFRs. The VDIC strategy is essentially the droop control \eqref{droop_des} with a time-variant droop coefficient. Hence, the key point of the VDIC design is to derive the expression of the droop coefficient based on the equivalent-scenario-based approach. To make the FFRs provide the required constant equivalent inertia support, we let the equivalent inertia coefficient $\Delta T_J$ be a constant and the droop coefficient $k^r$ be a variable. Since $\Delta T_J$ is a constant, from the first equality in \eqref{droop_rela}, we have:
\begin{align}
\label{omega_2}
\omega_e (t) = \frac{\Delta P_f}{T_J+\Delta T_J} t
\end{align} 
Substitute \eqref{omega_2} into the second equality in \eqref{droop_rela}, we can derive the time-domain expression of the droop coefficient of the VDIC strategy:
\begin{align}
\label{droop_exp}
k^r(t) = \frac{\Delta T_J}{t}
\end{align}

\textbf{Remark 2:} According to \eqref{droop_exp}, if the droop coefficient changes in inverse proportion to the post-fault time, the droop control of the FFRs will provide constant equivalent inertia support $\Delta T_J$. Therefore, the VDIC strategy with \eqref{droop_exp} can provide the required constant inertia support without introducing generator-like control dynamics, and thus avoids causing system oscillations. Moreover, the expression of $k^r(t)$ is independent of the power imbalance $\Delta P_f$. This is a fine property implying that the designed VDIC strategy can be set in advance and can deal with various power imbalances.

\begin{figure}[b]
	\centering
	\vspace{-0.1cm}  
	\setlength{\abovecaptionskip}{-0.05cm}   
	\includegraphics[width=0.3\textwidth]{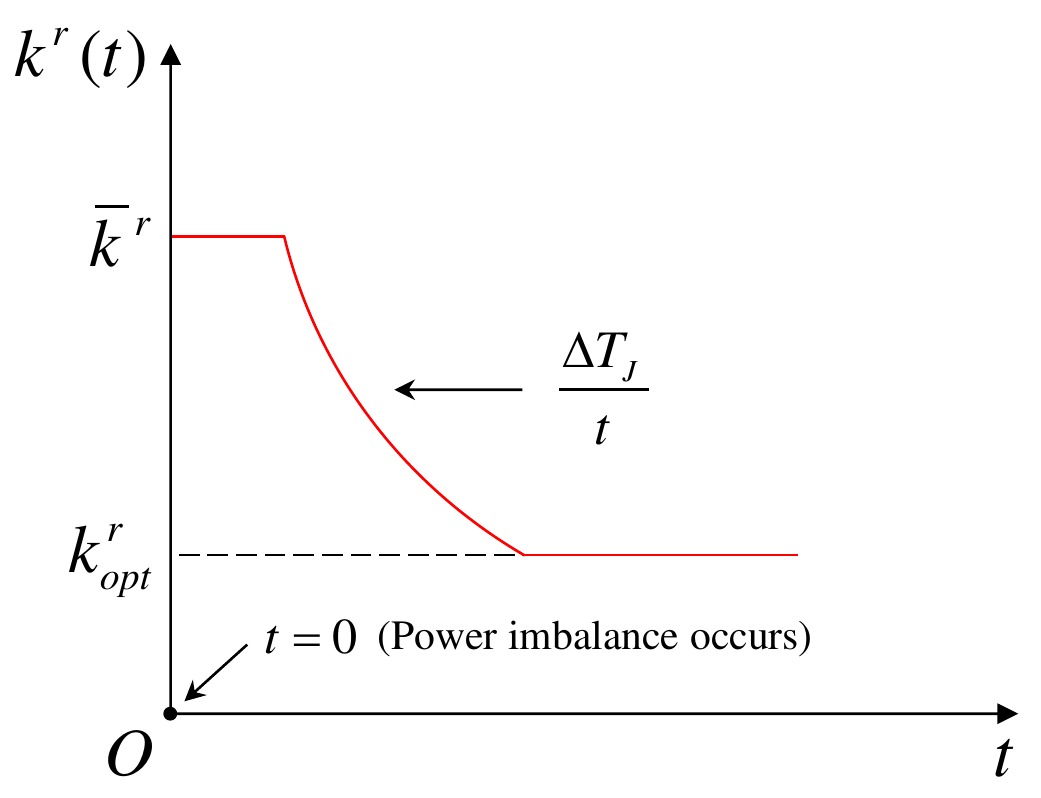}
	\caption{Diagram of the time-invariant droop coefficient.}
	\label{droop_coe}
\end{figure}

The designed VDIC strategy needs modifications for practical engineering applications. First, according to \eqref{droop_exp}, when $t \rightarrow 0$, $k^r(t) \rightarrow +\infty$; thus, considering the physical constraints of the FFRs in practice, the droop coefficient $k^r(t)$ needs a upper bound $\overline{k}^r$. Second, when $t \rightarrow +\infty$, $k^r(t) \rightarrow 0$. The droop control will lose the steady-state frequency support capability if $k^r(t) \rightarrow 0$. Hence, to give consideration to both the transient inertia support and the steady-state frequency support of the VDIC strategy, the droop coefficient $k^r(t)$ also needs a lower bound $\underline{k}^r$. The lower bound is usually set as the optimal constant droop coefficient $k^r_{opt}$ corresponding to a steady-state optimal control objective, e.g., the optimal droop coefficient in \cite{liu2021optimal} for power imbalance allocation. Moreover, since $k^r(t)$ is the total droop coefficient of multiple FFRs, in practical engineering, we need to appropriately allocate $k^r(t)$ among the FFRs, e.g., allocating $k^r(t)$ in proportion to the power regulation margins of the FFRs. With the above modifications, the time-domain diagram of the droop coefficient of the modified VDIC strategy is shown in Fig. \ref{droop_coe} (red line).

\section{Case Study}

In our previous work \cite{liu2021optimal}, a P-f droop control strategy is designed for conventional HVDC systems to provide fast emergency power support. Hence, in this section, we select the HVDC systems as the FFRs to verify the effectiveness of the VDIC strategy. In fact, the VDIC strategy can be applied to all types of FFRs. The test system is a hybrid AC-DC system where four HVDC systems are connected to an IEEE New England system. The electromagnetic transient (EMT) model of the test system is built on the CloudPSS platform \cite{liu2018modeling}. For detailed topology of the test system, we refer to \cite{liu2021optimal}. 

\begin{figure}[b]
	\centering
	\vspace{-0.2cm}  
	\setlength{\abovecaptionskip}{-0.05cm}   
	\includegraphics[width=0.42\textwidth]{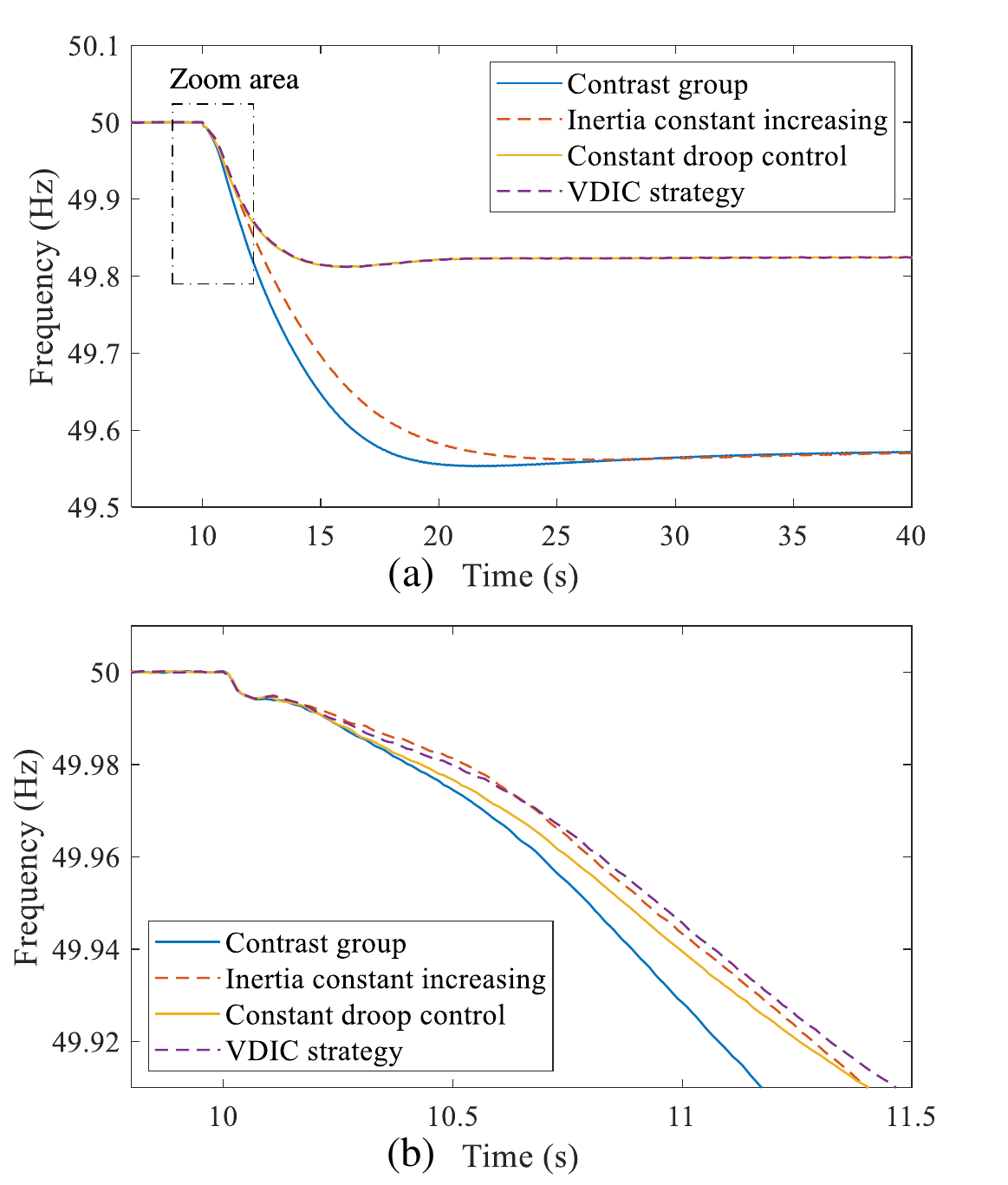}
	\caption{(a) System frequencies. (b) Zoom.}
	\label{case1}
\end{figure}

In this test system, the HVDC systems are implemented with the droop control, while the generators are implemented with relatively slow primary frequency control. We set the power base-value of the test system as $S_{sys} = 1000$ MVA. The actual inertia constant of the system is $T_J = 39.2$ s, and the required equivalent inertia support is assumed to be $\Delta T_J = 40$ s. We suppose that the upper bound and optimal value of the droop coefficient of each HVDC are 32 p.u. and 8 p.u.; thus, the upper bound and optimal value of the total droop coefficient are $\overline{k}^r = 128$ p.u. and $k^r_{opt} = 32$ p.u.. We set a power imbalance $\Delta P_f = -300$ MW at the time of 10 s. Based on the above settings, four subcases are simulated to verify the effectiveness of the VDIC strategy: (i) The actual system inertia constant remains unchanged and the droop control of the HVDCs is disabled (contrast group). (ii) The actual system inertia constant increases by $\Delta T_J$. (iii) The HVDCs adopt the droop control with the constant droop coefficient $k^r_{opt}$. (iv) The HVDCs adopt the VDIC strategy with the time-invariant droop coefficient in Fig. \ref{droop_coe}. The simulation results of the system frequencies in the above subcases are shown in Fig. \ref{case1}.      

Since the RoCoFs are negative in a period of time after the power imbalance $-300$ MW, in the following analysis, the ``RoCoF'' refers to the ``absolute value of the RoCoF'', i.e., the frequency with a larger RoCoF changes faster. First, according to Fig. \ref{case1}, within approximately 0.5 s after the power imbalance, the frequency changes in subcase (i) and (iii) are nearly identical; later, compared to the RoCoF in subcase (i), the RoCoF in subcase (iii) gradually decreases, which shows that the constant droop control of the HVDCs provides time-variant equivalent inertia support, and this inertia support increases gradually over time. Second, within approximately 1 s after the power imbalance, the frequencies in subcase (ii) and (iv) present nearly identical tendencies, and the RoCoFs in these two subcases are smaller than that in subcase (iii), which illustrates that the VDIC strategy has almost the same effect as actually increasing the system inertia constant in the post-fault initial period. Moreover, the steady-state frequency in subcase (iv) is closer to the nominal value than that in subcase (ii); thus, the VDIC strategy can provide both the transient equivalent inertia support and the steady-state frequency support. The above analysis verifies the effectiveness of the VDIC strategy.  

\section{Conclusion}
     
In this letter, we quantitatively investigate the equivalent inertia provided by the droop control of FFRs. First, an equivalent-scenario-based method is proposed to evaluate the equivalent inertia provided by the droop control, which shows that the droop control with a constant droop coefficient provides time-variant equivalent inertia support. Second, we design the VDIC strategy with a time-variant droop coefficient for the FFRs. The VDIC strategy can provide the required constant equivalent inertia support without introducing generator-like dynamics, and thus avoids causing system oscillations. In the case study where the HVDC systems are selected as the FFRs, the effectiveness of the proposed VDIC strategy is verified. 

\bibliographystyle{IEEEtran}
\bibliography{mybib}

\begin{thebibliography}{1}
\providecommand{\url}[1]{#1}
\csname url@samestyle\endcsname
\providecommand{\newblock}{\relax}
\providecommand{\bibinfo}[2]{#2}
\providecommand{\BIBentrySTDinterwordspacing}{\spaceskip=0pt\relax}
\providecommand{\BIBentryALTinterwordstretchfactor}{4}
\providecommand{\BIBentryALTinterwordspacing}{\spaceskip=\fontdimen2\font plus
\BIBentryALTinterwordstretchfactor\fontdimen3\font minus
  \fontdimen4\font\relax}
\providecommand{\BIBforeignlanguage}[2]{{%
\expandafter\ifx\csname l@#1\endcsname\relax
\typeout{** WARNING: IEEEtran.bst: No hyphenation pattern has been}%
\typeout{** loaded for the language `#1'. Using the pattern for}%
\typeout{** the default language instead.}%
\else
\language=\csname l@#1\endcsname
\fi
#2}}
\providecommand{\BIBdecl}{\relax}
\BIBdecl

\bibitem{milano2018foundations}
F.~Milano, F.~D{\"o}rfler, G.~Hug \emph{et~al.}, ``Foundations and challenges
  of low-inertia systems,'' in \emph{2018 power systems computation conference
  (PSCC)}.\hskip 1em plus 0.5em minus 0.4em\relax IEEE, 2018, pp. 1--25.

\bibitem{chen2020modelling}
M.~Chen, D.~Zhou, and F.~Blaabjerg, ``Modelling, implementation, and assessment
  of virtual synchronous generator in power systems,'' \emph{J. Mod. Power
  Syst. Clean Energy}, vol.~8, no.~3, pp. 399--411, 2020.

\bibitem{shang2021fast}
L.~Shang, X.~Dong, C.~Liu \emph{et~al.}, ``Fast grid frequency and voltage
  control of battery energy storage system based on the
  amplitude-phase-locked-loop,'' \emph{IEEE Trans. Smart Grid}, vol.~13, no.~2,
  pp. 941--953, 2022.

\bibitem{du2019power}
W.~Du, Q.~Fu, and H.~Wang, ``Power system small-signal angular stability
  affected by virtual synchronous generators,'' \emph{IEEE Trans. Power Syst.},
  vol.~34, no.~4, pp. 3209--3219, 2019.

\bibitem{meng2018generalized}
X.~Meng, J.~Liu, and Z.~Liu, ``A generalized droop control for grid-supporting
  inverter based on comparison between traditional droop control and virtual
  synchronous generator control,'' \emph{IEEE Trans. Power Electron.}, vol.~34,
  no.~6, pp. 5416--5438, 2019.

\bibitem{van2015droop}
J.~Van~de Vyver, J.~D. De~Kooning, B.~Meersman \emph{et~al.}, ``Droop control
  as an alternative inertial response strategy for the synthetic inertia on
  wind turbines,'' \emph{IEEE Trans. Power Syst.}, vol.~31, no.~2, pp.
  1129--1138, 2016.

\bibitem{liu2021optimal}
Y.~Liu, Y.~Song, Z.~Wang \emph{et~al.}, ``Optimal emergency frequency control
  based on coordinated droop in multi-infeed hybrid {AC}-{DC} system,''
  \emph{IEEE Trans. Power Syst.}, vol.~36, no.~4, pp. 3305--3316, 2021.

\bibitem{liu2018modeling}
Y.~Liu, Y.~Song, Z.~Yu \emph{et~al.}, ``Modeling and simulation of hybrid
  {AC-DC} system on a cloud computing based simulation platform-{C}loud{PSS},''
  in \emph{2018 2nd IEEE Conference on Energy Internet and Energy System
  Integration (EI2)}.\hskip 1em plus 0.5em minus 0.4em\relax IEEE, 2018, pp.
  1--6.

\end{thebibliography}

\end{document}